\documentstyle [12pt] {article}

\newcommand{\mfrac}[2]{\displaystyle \frac{\displaystyle #1}{\displaystyle #2}}
\newcommand{\cs}[3]{{{#3} \brace {#1 #2}}}

\newcommand{\h}[1]{\mathop{\lambda}\limits_{#1}\ \!\!\!}

\newcommand{\edf}{\ {\mathop{=}\limits^{\rm def}}\ }
\newcommand{\half}{\textstyle \frac{1}{2} \displaystyle}

\newcommand{\Lag}{{\cal L}}
\newcommand{\al}{\alpha}
\newcommand{\be}{\beta}
\newcommand{\m}{\mu}
\newcommand{\n}{\nu}
\newcommand{\s}{\sigma}
\newcommand{\g}{\gamma}

\newcommand{\de}{\delta}

\begin{document}
\newpage
\title{\bf " Hamilton's principle and the Generalized
Field Theory : A comparative study "}

\author{{\bf F. I. Mikhail}\\
\normalsize  Department of mathematics, faculty of science,
Ain shames university, {\bf Egypt} \and
{\bf M. I. Wanas}\\
\normalsize Astronomy Department, Faculty of Science,
Cairo university {\bf Egypt}}

\newpage
\maketitle
\begin{abstract}
The field equations of the generalized field theory (GFT) are derived from
an action principle. A comparison between (GFT), M\o ller's tetrad theory of
gravitation (MTT), and general relativity is carried out regarding the
Lagrangian of each theory. The results of solutions of the field equations,
of each theory, are compared in case of spherical symmetry. The differences
between the results are discussed and interpreted.
\end{abstract}

\newpage

\section{Introduction:}

   In the last fifteen years several field theories appeared in literature,
   each of which is assumed to be a natural generalization of the theory
   of general relativity. It is high time to carry out a thorough comparison
   between these theories with regard to three fundamental pivots of any
   generalized field theory, i.e:
\begin{description}
   \item{(i)} the geometric structure used in the formalism of the theory,
   \item{(ii)} the procedure used in the derivation of its field equations,
   \item{(iii)} the models to which these field equations are applied and 
the results, of physical interest, obtained.
\end{description}

   We believe that such comparison will be more feasible if we are able
   to unify these three aspects in the theories under consideration.

   The present work is a trail to carry out such a comparison along the
   lines specified above. Strictly speaking, we are going to examine the
   two generalized  field theories :
\begin{description}
\item{(a)} the generalized field theory (GFT) constructed by Mikhail \& Wanas
   (1977),
\item{(b)} M\o ller's tetrad theory of gravitation (MTT) (1978),
\end{description}
   versus two different versions of general relativity (GR): the standard
   orthodox theory in which the lorentz signature is being imposed on the
   metric used from the beginning, and a modified version given by Wanas
   (1990) in which the Lorentz signature is introduced at a later stage.

        Fortunately, the two theories (a), (b) under consideration depend
   on a Absolute Parallelism (Ap-Space) in their formalism. Thus the pivot$(i)$
   mentioned above is already unified. To satisfy pivot$(ii)$ we have to
   derive the field equations of (GFT) using a variational principle, i.e.
   Hamilton's action principle. This is given in section 4 below. Besides,
   for comparison with (GR), we are going to use a Lagrangian function given
   by M\o ller (1978) as shown in section 6.

  With regards to pivot $(iii)$, we are going to apply the two
  theories (a),(b) and both versions of (GR) to the same absolute parallelism
  (AP) space, namely the most general AP-Space having spherical symmetry
  derived by Robertson (1932). The results obtained by applying the four
  different theories to this same AP-space are compared and tabulated in
  section 7.

\setcounter{equation}{0}
\section{Basic Geometric Structure (AP-space)}

  We are mainly interested in field theories using for their
  structure a geometry admitting absolute parallelism (AP-geometry).
  The AP-space is a 4-dimensional vector space $(T_4)$, each point of which is
  labelled by 4-independent variables $x^\nu (\nu=0,1,2,3)$. At each point,
  4-linearly independent contravariant vectors $\h{i}^\m , (i=0,1,2,3)$,
  are defined. Assuming that the $\h{i}_\m$, is the
  normalized cofactor of $\h{i}^\m$ in the determinant $||\h{i}^\m||$.
  Hence they satisfy the relations,

\begin{equation}
\h{i}^\m \h{j}_\m=\de_{ij}
\end{equation}

\begin{equation}
 \h{i}^\m \h{i}_\n=\de^\m_{\ \n}
 \end{equation}
In what follows we use Greek letters $(\m,\n,\al,\be,...)$ to indicate
coordinate components (world indices), and Latin letters $(i,k,...)$ to
indicate vector numbers (mesh indices). Summation convention will be applied to
both types of indices. Using these vectors we can define the following
2\underline {nd} order tensor:
\begin{equation}
\hat{g}_{\n\m} \edf \h{i}_\n \h{i}_\m ,
\end{equation}

\begin{equation}
\hat{g}^{\n\m} \edf \h{i}^\n \h{i}^\m .
\end{equation}
It can be shown that
\begin{equation}
\hat{g}^{\m\al}g_{\n\al}\edf \de^\m_{\ \n} ,
\end{equation}
\begin{equation}
\hat{g}=\hat{\lambda}^2 ,
\end{equation}
where $\hat{g}\edf ||\hat{g}_{\m\n}||$, and $\hat{\lambda}\edf ||\h{i}_\m||$.\\
we can use $\hat{g}^{\m\n}$ and $\hat{g}_{\al\be}$ to raise and lower world indices.
It can also be used as a metric tensor defining a Riemannian space
associated with the AP-space when needed. A non symmetric connection
$\Gamma^\al_{\m\n}$ is then defined such that,
\begin{equation}
\h{i}_{{\stackrel{\m}{+}} | \n}\edf
 \h{i}_{\m ,\n}-\Gamma^\al_{\ \m\n} \h{i}_\al =0 .
\end{equation}
 This is the condition of absolute parallelism. Equation (2.7) can directly be
 solved to give,

\begin{equation}
\Gamma^\al_{\ \m\n}=\h{i}^\al\h{i}_{\m,\n} .
\end{equation}

It can be shown that (cf.Mikhail (1962)):
\begin{equation}
 \Gamma^\al_{\ \m\n}=\cs{\m}{\n}{\al}+\gamma^\al_{\ \m\n} ,
\end{equation}
where $\cs{\m}{\n}{\al}$\,is Christoffel symbols defined as usual using the
tensors (2.3), (2.4), and\, $\gamma^\al_{\ \m\n}$ is a 3rd order tensor defined
by,
\begin{equation}
\g^\al_{\ \m\n}\edf\h{i}^\al\h{i}_{\m ;\n} .
\end{equation}
So,in the AP-space, we have three different connexions:\,$\cs{\m}{\n}{\al}$,
$\Gamma^\al_{\ \m\n}$ and $\Gamma^\al_{\ (\m\n)}(=\half(\Gamma^\al_{\ \m\n}+\Gamma
^\al_{\ \n\m}))$ .Consequently, we can define the following types of absolute
derivatives:
\begin{equation}
 A^\m_{\ ;\n} \edf A^\m_{\ ,\n} + \cs{\n}{\al}{\m} A^\al ,
\end{equation}

\begin{equation}
 A^{{\stackrel{\m}{+}}}_{\ | \n}\edf A^{\m}_{\ ,\n} + \Gamma^{\m}_{\ \al_\n} A^\al ,
\end{equation}
\begin{equation}
 A^{{\stackrel{\m}{-}}}_{\ | \n}\edf A^{\m}_{\ ,\n}+\Gamma^{\m}_{\ \n \al} A^\al ,
\end{equation}
\begin{equation}
 A^{\m}_{\ | \n} \edf A^{\m}_{\ ,\n}+\Gamma^{\m}_{\ (\al \n)} A^\al ,
\end{equation}
where $A^\m$ is an arbitrary contravariant vector and the comma (,) denotes
 ordinary partial differentiation.

The torsion tensor is defined by
\begin{equation}
\Lambda^{\al}_{\ \m\n}\edf\Gamma^{\al}_{\ \m\n}-\Gamma^{\al}_{\ \n\m}=-\Lambda^{\al}_{\ \n\m} .
\end{equation}
Contracting this tensor by setting $\n=\al$ ,we get the basic vector
\begin{equation}
C_{\m}\edf\Lambda^\al_{\ \m\al}=\gamma^\al_{\ \m\al} .
\end{equation}
For more details see Mikhail (1952).

\setcounter{equation}{0}
\section{Field Equations of (GFT):}

Using the AP-space, as described in the above section, Mikhail \& Wanas (1977)
were able to construct a generalized field theory (GFT). Using a certain Lagrangian
function, they were able to derive an identity of the form,
\begin{equation}
E^{{\stackrel{\m}{+}}}_{\ {\stackrel{\n}{+}}|\m}=0 .
\end{equation}
Following some analogy with (GR), they were led to chose their field equations
into the form

\begin{equation}
E^\m_{\ \n}=0,
\end{equation}

where $E^\m_{\ \n}$ is a non-symmetric tensor given by:
\begin{eqnarray}
E^{\m}_{\ \n} & \edf & {\de}^{\m}_{\ \n} L-2L^{\m}_{\ \n}
- 2{\de}^{\m}_{\ \n} C^{\stackrel{\al}{-}}_{\ |\al}\nonumber \\
& &  -2 C^{\mu} C_{\nu}
-2 {\de}^{\m}_{\ \al} C^{\be} \Lambda^{\al}_{\ \be\n}
+2\hat{g}^{\al \m} C_{{\stackrel{\n}{+}} | \al}
-2\hat{g}^{\al\be} {\de}^{\m}_{\ \epsilon} \Lambda^{{\stackrel{\epsilon}{+}}}
_{\ \ {\stackrel{\nu}{+}}{\stackrel{\be}{+}} | \al} ,
\end{eqnarray}
and  ${L}_{\m\n}$ is given by
\begin{eqnarray}
{L}_{\m \n} & \edf & \Lambda^{\al}_{\ \be\m}\,
\Lambda^{\be}_{\ \al\n}-C_{\m}C_{\n} \nonumber, \\
 L &\edf& \hat{g}^{\m \n}L_{\m \n}.
\end{eqnarray}
Several promising solutions of this theory have been obtained,e.g. 
Wanas (1985, 1987, 1989).

However, it is of great interest to find out the particular lagrangian function
which will give rise to the field equations (3.2) by using a variational
principle. This is what we intend to do in the following section. In fact we were
able to find out the Lagrangian function L giving rise to the scalar density,

$${\cal L}\edf{\hat \lambda}L ,$$
and then using the variational principle,
$$ {\de}\int{\cal L}\ d^{4}x =0 , $$
we will try to derive the field equations (3.2).

\setcounter{equation}{0}

\section{Derivation of the field equation of GFT using Hamilton's
action principle}

Generally speaking, starting with a Lagrangian function L which
depends only on the tetrad vectors and their first coordinate derivatives
i.e.
\begin{equation}
 L=L(\h{i}_{\m},\h{i}_{\m,\n}),
\end{equation}
we can define the scalar density,
\begin{equation}
{\cal L}\edf{\hat \lambda}L ,
\end{equation}
and thus :
\begin{equation}
 {\cal L}={\cal L}(\h{i}_{\m},\h{i}_{\m,\n}).
\end{equation}
Then the action principle will give
\begin{equation}
 \de \int{\cal L}(\h{i}_{\m},\h{i}_{\m,\n})\ d^{4}x =0.
\end{equation}
where $d^4x=dx^0 dx^1 dx^2 dx^3 $ .
Then considering an arbitrary variation $(\de\h{i}_{\m})$ in the
tetrad vectors, (4.4) can be put in the form,
\begin{equation}
\int\frac{\de{\cal L}}{\de\h{i}_\m}\,\de\h{i}_{\m}\ d^{4}x=0,
\end{equation}

where 
\begin{equation}
 \frac{\de{\cal L}}{\de\h{i}_\m}\edf \frac{\partial {\cal L}}{\partial \h{i}_{\m}}
-{\frac{\partial}{\partial x^\g}}
\left({\frac{\partial {\cal L}} {\partial \h{i}_{\m ,\g}}}\right),
\end{equation}
is the Hamiltonian derivative of ${\cal L}$ w.r.t. the arbitrary variation
$(\de\h{i}_\m)$. On the other hand we can write the Hamiltonian derivative
(4.6) in the form,
\begin{equation}
\frac{\de{\cal L}}{\de\h{i}_\m}\edf\hat{\lambda} S_{i}^\m.
\end{equation}
It can be easily shown that $S_{i}^\m$ is a vector for each value of $i$
(Wanas 1975). Using $S_{i}^\m$, we can define the following $2$\underline{nd}
order tensor,
\begin{equation}
S^{\be}_{\ \s}\edf\h{i}_{\s} S_{i}^\be .
\end{equation}
We assume that $(\de\h{i}_\be)$ are linearly independent, and arbitrary as
stated before, then for (4.5) to be satisfied, we write the Euler-Lagrange
equation for the present problem :

\begin{equation}
 \frac{\de{\cal L}}{\de\h{i}_\m}=
\frac{\partial {\cal L}}
       {\partial \h{i}_{\be}}
-{\frac{\partial}{\partial x^\al}}
\left({\frac{\partial {\cal L}} {\partial \h{i}_{\be ,\al}}}\right)=0.
\end{equation}
Then using the same Lagrangian function defined by (3.4) in (GFT), the
1{\underline {st}} term of (4.9) will take, after some manipulation,
the form :

\begin{eqnarray}
\frac{\partial {\cal L}}{\partial \h{i}_\be} & = & \hat{\lambda}\,\h{i}^{\be}
\,\hat{g}^{\m\n}\,L_{\m\n}-\hat{\lambda}\left(\hat{g}^{\m\be}\h{i}^{\n}
+\hat{g}^{\n\be}\h{i}^{\m}\right) L_{\m\n}\nonumber \\
& & +\hat{\lambda}\hat{g}^{\m\n}
\left(\h{i}^{\al}\Lambda^{\be}_{\ \m\epsilon}\Lambda^{\epsilon}_{\ \al\n}
+\h{i}^{\epsilon}\Lambda^{\al}_{\ \epsilon\m}\Lambda^{\be}_{\ \n\al}
-\h{i}^{\al}\Lambda^{\be}_{\ \al\m}C_{\n}
-\h{i}^{\al}\Lambda^{\be}_{\ \al\n}C_{\m}\right).
\end{eqnarray}

Similarly, we get for the $2{\underline{nd}}$ term of (4.9) the expression :

\begin{eqnarray}
\frac{\partial}{\partial x^\g}\,\frac{\partial\Lag}{\partial \h{i}_{\be,\g}} &=& 
2\hat{\lambda}\,\Gamma^{\m}_{\ \m\g}
\left( \h{i}^{\epsilon}\hat{g}^{\g\al}\Lambda^{\be}_{\ \epsilon\al}
-\h{i}^{\epsilon} \hat{g}^{\al\be}\Lambda^{\g}_{\ \epsilon\al}
-\h{i}^{\g} \hat{g}^{\al\be}C_{\al}+\h{i}^{\be} \hat{g}^{\g\al} C_{\al} \right)\nonumber \\
& & +2\hat{\lambda}\left( \h{i}^{\epsilon}_{\ ,\g}\right.
\hat{g}^{\g\al}\Lambda^{\be}_{\ \epsilon\al}
+\h{i}^{\epsilon} \hat{g}^{\g\al}_{\ \ ,\g} \Lambda^{\be}_{\ \epsilon\al}
+\h{i}^{\epsilon} \hat{g}^{\g\al}\Lambda^{\be}_{\ \epsilon\al,\g}\nonumber \\
& &  - \h{i}^{\epsilon}_{\ ,\g} \hat{g}^{\al\be} 
 \Lambda^{\g}_{\ \epsilon\al} 
 -\h{i}^{\epsilon} \hat{g}^{\al\be}_{\ \ ,\g}\Lambda^{\g}_{\ \epsilon\al} 
 -\h{i}^{\epsilon} \hat{g}^{\al\be}\Lambda^{\g}_{\ \epsilon\al,\g}\nonumber \\ 
& & -\h{i}^{\g}_{,\g} C^{\be} 
- \h{i}^{\g} C^{\be}_{\ ,\g} - \h{i}^{\be}_{\ ,\g}C^{\g}
\left.+\h{i}^{\be} C^{\g}_{\ ,\g}\right).\ \ \ \ \ \ \ \
\end{eqnarray}
Substituting from (4.10), (4.11) into Euler-Lagrange equation (4.9) we get,
after comparing results with (4.7) \& (4.8) the following set of field
equations, (since $\hat{\lambda}\neq 0$ and $\h{i}_\s$ are linearly
independent)
\begin{equation} 
S^\be_{\ \s}=0,
\end{equation}
where

\begin{eqnarray}
S^\be_{\ \s} & \edf & \de^\be_{\ \s}L -2 L^\be_{\ \s}
+2\hat{g}^{\m\al}\Lambda^{\be}_{\ \m\n}\left(
\de^\epsilon_{\ \s}\Lambda^{\n}
_{\ \epsilon\al}+\de^\n_{\ \s} C_\al \right)\nonumber \\
 & & -2\left[\Gamma^\m_{\ \m\g}\right.\left(\de^\epsilon_{\ \s}\hat{g}^{\g\al}
\Lambda^\be_{\ \epsilon\al}
- \de^{\epsilon}_{\ \s}\hat{g}^{\al\be}\Lambda^{\g}
_{\ \epsilon\al} - \de^\g_{\ \s}\hat{g}^{\al\be}C_\al
+ \de^\be_{\ \s} \hat{g}^{\g\al}C_\al\right)\nonumber \\
& &  -\Gamma^\epsilon_{\ \s\g} \hat{g}^{\g\al}\Lambda^{\be}_{\ \epsilon\al}
+\de^\epsilon_{\ \s}\hat{g}^{\g\al}_{\ \ ,\g}\Lambda^{\be}_{\ \epsilon\al}
+\de^\epsilon_{\ \s} \hat{g}^{\g\al}\Lambda^{\be}_{\ \epsilon\al ,\g}
 -\Gamma^\epsilon_{\ \s\g}\hat{g}^{\al\be}\Lambda^{\g}_{\ \epsilon\al}\nonumber \\
& &-\de^\epsilon_{\ \s}\hat{g}^{\al\be}_{\ \ ,\g}\Lambda^{\g}_{\ \epsilon\al}
-\de^\epsilon_{\ \s}\hat{g}^{\al\be}\Lambda^{\g}_{\ \epsilon\al ,\g}
+\Gamma^\g_{\ \s\g} C^\be -\de^\g_{\ \s} C^\be_{\ ,\g}
-\Gamma^\be_{\ \s\g}C^\al
    +\left. \de^\be_{\ \s} C^\g_{\ ,\g}\right].\ \ \ \ \ \ \ \ \
\end{eqnarray}
Using the following results (Wanas ((1975))
$$ C^{\stackrel{\n}{-}}_{\ |\s}-C^{\stackrel{\n}{+}}_{\ |\s}=-C^\epsilon
\Lambda^{\n}_{\ \epsilon\s},$$

$$\hat{g}^{{\stackrel{\g}{+}}{\stackrel{\al}{-}}}_{\ \ \ |\g}=0\ \ \ 
,\ \ \  \hat{g}^{{\stackrel{\al}{+}}{\stackrel{\n}{+}}}_{\ \ \ |\s}=0,\ \  $$
and the identity (Einstein (1929))
$$ \Lambda^{\stackrel{\g}{-}}_{\ \ {\stackrel{\s}{+}}{\stackrel{\al}{+}}|\g}
=C_{{\stackrel{\s}{+}}|\al}-C_{{\stackrel{\al}{+}}|\s}, $$
then (4.13) can be written in the form,

then (4.13) can be written in the form
\begin{eqnarray}
S^{\be}_{\ \s} & \edf & {\de}^{\be}_{\ \s} L-2L^{\be}_{\ \s}
- 2{\de}^{\be}_{\ \s} C^{\stackrel{\g}{-}}_{\ |\g}\nonumber \\
& &  -2 C^{\be} C_{\s}
-2 {\de}^{\be}_{\ \n} C^{\epsilon} \Lambda^{\n}_{\ \epsilon\s}
+2\hat{g}^{\al \be} C_{{\stackrel{\s}{+}} | \al}
-2\hat{g}^{\g\al} {\de}^{\be}_{\ \n} \Lambda^{{\stackrel{\n}{+}}}
_{\ \ {\stackrel{\s}{+}}{\stackrel{\al}{+}} | \g} ,
\end{eqnarray}
which is identical with the tensor $E^\m_{\ \n}$ given by (3.3) in the Mikhail-
Wanas derivation. This shows that the field equations of (GFT) can be derived
from an action principle with the Lagrangian function given by (3.4), (4.2).

\setcounter{equation}{0}
\section{Role of the signature}
  In metric theories of gravity, the Lorentz signature plays a very
  important role. The signature is defined as the difference between the
  number of +ve and -ve eigenvalues of the metric tensor. Lorentz signature
  is imposed on the metric, for purely physical reasons, namely to account
  for the special relativity in the limiting case. This reflects the fact
  that we distinguish between spatial and temporal sections of space-time. It
  was realized, since the appearence of special relativity, that our
  universe is
  a 4-dimensional one. So, there is no fundamental need to distinguish between
  space and time from the beginning. We are unable neither, to construct
  equipment or experiment, nor to measure in 4-dimensions, but rather in (3+1)
  dimensions. Thus, in order to compare the results of field theories with
  observation and experiment, Lorentz signature should be introduced at a later
  stage in the theory to reflect our distinction between space \& time.
  In fact, the introduction  of Lorentz signature transfers the theory from
  4-dimensions to (3+1) dimensions.

  Wanas (1990) suggested that the insertion of Lorentz signature is to be done
  just before matching the results of the theory with observations or
  experiments, i.e. after solving the field equations. He speculated that
  this may give rise to new physics, and give an example confirming his
  speculation.

     There are different ways for inserting the Lorentz signature in a theory
     constructed in the AP-space. One way is just to change the sign of some
     constants of integration after solving the field equations. Another way
     is to replace (2.3) by
     $$g_{\m\n}\edf e_i\h{i}_\m \h{i}_\n$$
     where $e_i=(\it{+1,-1,-1,-1})$\ is the Levi-Civita indicator.
     A 3\underline{rd} way is to take the vector $\h{\it{0}}^\m$ to be imaginary.
      For AP-space whose associated metric is diagonal
      (or it could be diagonalized), it doesn't matter whether we insert the
      Lorentz signature before or after solving the field equations.

\section{Direct comparison between the four theories :}

    So far we have satisfied the two factors $(i)$ \& $(ii)$
  mentioned in the introduction. It may be of interest to compare
  the four theories, under consideration, with regard to the
  Lagrangian function used and the field equations derived in each.

    For standard (GR) (written in the AP-space ) the Lagrangian is usually
    taken to be (cf. M\o ller 1978),
\begin{equation}
{\cal L}_{GR} = \sqrt{-g}\,({\g}^{\al\m\n}{\g}_{\n\m\al}-C^{\n}C_{\n}).
\end{equation}
Consequently, the modified version of (GR) as speculated by Wanas (1990) will
be,
\begin{equation}
{\cal L}_{\hat{GR}} =\sqrt{\hat{g}}\,({\g}^{\al\m\n}{\g}_{\n\m\al}-C^{\n}C_{\n}).
\end{equation}
The corresponding Lagrangian of (GFT), as given above,
\begin{equation}
{\cal L}_{GFT}=
\sqrt{\hat{g}}\,(3{\g}^{\al\m\n}{\g}_{\n\m\al}
-{\g}^{\al\m\n}{\g}_{\al\m\n}-C^{\n}C_{\n}).
\end{equation}
M\o ller (1978) has chosen for his theory the Lagrangian function,
\begin{equation}
{\cal L}_{MTT}= \sqrt{-g}\,((1-2{\psi}){\g}^{\al\m\n}{\g}_{\n\m\al}
+{\psi}{\g}^{\al\m\n}{\g}_{\al\m\n}-C^{\n}C_{\n}),
\end{equation}
where $\psi$ is a free parameter of order unity. It is clear that ${\cal L}_{GR}$
,${\cal L}_{GFT}$ can be obtained as special cases of M\o ller's
Lagrangian (5.3) by taking,

$$ \psi =0\ \ \ \ \ \ \ \ \ ,\ \ \ \ \ \ \ \ \ \ \ \ \ \psi =-1,$$
respectively. However, although the lagrangian used by M\o ller in (MTT)
appears to be more general than those for deriving the field equations
of (GR), and (GFT), yet M\o ller's theory (MTT) as a whole is no so general.
This will be shown clearly in the next section.

  The reason, in our opinion, is due to the following two factors :
  \begin{description}
\item [{\it(i)}]  In M\o ller's theory, the Lorentz signature is imposed on the theory
  from the beginning.
\item[{\it(ii)}] M\o ller used a phenomenological definition for the material-energy tensor
  $T_{\m\n}$, while it is defined geometrically in (GFT).
  \end{description}
Formally, (6.1) \& (6.2) are similar, but the results of application are
different. We are going to consider the theory depending on (6.2) as different
compared to (GR). The following table summarizes the main features of each theory.

\small

\begin{center}
Table 1: Comparison between the four field theories
\vspace{1.3cm}
\begin{tabular}{|c|c|c|c|c|c|} \hline
& & & & & \\
Field & Field & Field & Field & Gravitational & $T_{\mu\nu}$ \\
Theory & Lagrangian &  Equations & Variables & Potential & \\
& & & & & \\
\hline
& & & & & \\
GR (1916) & ${\cal L}_{GR}$ & $G_{\mu\nu}=-\kappa T_{\mu\nu}$ & $g_{\mu\nu}$
& $g_{\mu\nu}$ & Phenom. \\
& & & & & \\  \hline
& & & & & \\
GFT (1977) & ${\cal L}_{GFT}$ & $\hat{G}_{\mu\nu} = B_{\mu\nu}$
& $\h{i}_\mu$ & $g_{\mu\nu}$ & Geomet. \\
& & $E_{[ \mu\nu ]}=0$ & & & \\
& & & & & \\ \hline
& & & & & \\
MTT (1978) & ${\cal L}_{MTT}$ & $G_{\mu\nu}+H_{\mu\nu}=-\kappa T_{\mu\nu}$
& $\h{i}_\mu$ & $g_{\mu\nu}$ & Phenom. \\
& & $V_{[ \mu\nu ]}=0$ & & & \\
& & & & & \\ \hline
& & & & & \\
$\hat{\hbox{GR}}$ (1990) & $\hat{{\cal L}}_{GR}$ & $\hat{G}_{\mu\nu}=-\kappa
T_{\mu\nu}$ & $\hat{g}_{\mu\nu}$ &  $g_{\mu\nu}$ & Phenom. \\
& & & & & \\ \hline
\end{tabular}
\end{center}
where $G_{\m\n}$ is Einstein tensor.

\section{Comparison in the case of spherical symmetry}

   Since all theories tabulated in table 1 are written in the AP-space, it is
 convenient to compare them by using the most general AP-space having
 spherical
 symmetry. The structure of this space is given by Robertson (1932). The tetrad
 giving the structure of this space can be written in spherical polar
 coordinates as,

\begin{equation}
\h{i}^ \mu=\left(\matrix{
A &  Dr  & 0 & 0 \cr\cr
0 & B\sin\theta\cos\phi & \mfrac{B}{r}\cos\theta\cos\phi
  & -\mfrac{B}{r}\mfrac{\sin\phi}{\sin\theta} \cr\cr
0 & B\sin\theta\sin\phi & \mfrac{B}{r}\cos\theta\sin\phi
  & \mfrac{B}{r}\mfrac{\cos\phi}{\sin\theta} \cr\cr
0 & B\cos\theta & -\mfrac{B}{r}\sin\theta & 0
}\right),
\end{equation}
where A,B,D are unknown functions of $r$ only. Calculating the necessary tensors
for this space we find that (see table 1)\,:

\begin{description}
\item[{\it (1)}] For M\o ller's tetrad theory $H_{\mu\nu}=0$\ ,\ $V_{[\mu\nu]}=0$ is
satisfied
identically.
\item [{\it (2)}] For GFT $B_{\mu\nu}=0$\ ,\ $E_{[\mu\nu]}=0$ is satisfied identically.
\end{description}
Furthermore, if we direct our attention to free space solutions $(T_{\m\n}=0)$
we found the results summarized in table 2.

\begin{center}
Table 2: Comparison in the case of spherically symmetric solutions
\vspace{1.3cm}
\begin{tabular}{|c|c|c|c|c|} \hline
& & & & \\
Field & Field & Schwarz. & Other & Reference \\
Theory & equations & solution? & Solutions ? &  \\
& & & & \\
 \hline
& & & & \\
GR & $G_{\mu\nu}=0$ & Yes & No & Birkhoff Theorem  \\
& & & & \\  \hline
& & & & \\
GFT & $\hat{G}_{\mu\nu}=0$ & Yes & Yes & Wanas $(1985)$ \\
& & & & \\ \hline
& & & & \\
MTT & $ G_{\mu\nu}=0 $ & Yes &  No & Mikhail et. al. $(1991)$ \\
& & & & \\  \hline
& & & & \\
$\hat{GR}$ & $\hat{G}_{\mu\nu}=0$ & Yes & Yes & Wanas $(1990)$ \\
& & & & \\  \hline
\end{tabular}
\end{center}

It is clear from the second column in table(2) above that although the
field equations of the four theories reduce formally to those of GR in free
space, the results obtained are not identical. It is clear that this
difference is a direct consequence to the stage at which  Lorentz
signature is introduced in the theory.
\newpage

{\bf\Large {References:}} \\
A. Einstein, {Sitz. der Preuss. Akad. der Wiss.} {\bf 1}, 2 (1929).\\
F. I. Mikhail, {\it Ph.D. Thesis}, University of London (1952).\\
F. I. Mikhail, {\it Ain Shams Sc. Bul.} {\bf 6}, 87 (1962).\\
F. I. Mikhail and M. I. Wanas, {\it Proc. Roy. Soc. Lond. A} {\bf 356},
471 (1977).\\
F. I. Mikhail and M. I. Wanas, {\it Int. J. Theoret. Phys.} {\bf 20},
671 (1981).\\
F. I. Mikhail, M. I. Wanas, A. A. Hindawi, E.I.Lashin
{\it Int. J. Theoret. Phys.} {\bf 32}, 1627 (1993). \\
C. M\o ller, {\it Mat. Fys. Skr. Dan. Vid. Selsk.} {\bf 39}, No.
13, 1 (1978). \\
H. P. Robertson, {\it Annals of Mathematics Princeton} {\bf 33}, 496 (1932). \\
M. I. Wanas, {\it Ph.D. Thesis}, Cairo University (1975). \\
M. I. Wanas, {\it Int. J. Theor. Phys.} {\bf 24}, 639 (1985). \\
M. I. Wanas, {\it ICTP Preprint} {\bf IC/87/39}. \\
M. I. Wanas, {\it Astrophys. Space Sci.} {\bf 154}, 165 (1989). \\
M. I. Wanas, {\it Astron. Nachr.} {\bf 311}, 253 (1990).
\end{document}